\ifx\pdfoutput\undefined
\else
  \pdfoutput=1 
\fi

\documentclass[aps,prb,reprint,amsmath,amssymb]{revtex4-2}

\usepackage[T1]{fontenc}
\usepackage[utf8]{inputenc} 
\usepackage{lmodern}        
\usepackage{amsmath,amssymb,bm}
\usepackage{mathtools}      
\usepackage{dcolumn}        
\usepackage{physics}        
\usepackage{xcolor}         
\usepackage{graphicx}       
\DeclareGraphicsExtensions{.pdf} 

\usepackage[caption=false]{subfig}

\usepackage{microtype}      
\usepackage{enumitem}       

\usepackage[colorlinks=true,allcolors=blue]{hyperref}

\newcommand{\Det}{\mathrm{Det}\,}
\newcommand{\Trr}{\Tr_r}
\newcommand{\Detr}{\mathrm{Det}_r\,}

\newif\ifdraft
\draftfalse            
\ifdraft
  
  \newcommand{\cin}[1]{\textcolor{red}{#1}}
\else
  \newcommand{\cin}[1]{#1}    
\fi

\begin{document}

\title{
Gradient flow  towards quantum states with ideal quantum geometry
}

\author{Takumi Shiga and Takahiro Fukui}
\affiliation{Department of Physics, Ibaraki University, Mito 310-8512, Japan}

\date{\today}

\begin{abstract}
We propose a gradient-flow method for quantum states in lattice models, generated
by an action consisting of the quantum metric  and the square of the Berry 
curvature. These two terms drive the spectral projector toward Bogomolny
saturation and uniform Berry curvature, respectively. We show that, due to a
no-go theorem for finite-dimensional projectors, the two conditions cannot in
general be satisfied simultaneously in lattice models. 
Hence the flow is expected to approach a nontrivial fixed point that balances the two geometric requirements. 
For the Wilson-Dirac model, we demonstrate that the flowed projector exhibits 
almost uniform Berry curvature while remaining close to the Bogomolny bound.
We further construct a short-range truncated flattened Hamiltonian from the
flowed projector and obtain a lattice model with nearly flat bands and nearly
uniform Berry curvature.
We also apply the method to the Hofstadter model and confirm
the roles of the metric and Berry-curvature terms in a Chern band with higher
Chern number.
\end{abstract}


\maketitle

\section{Introduction}

Topological quantum states are characterized not only by global topological
invariants, but also by local geometric properties of states
\cite{Provost:1980aa}. For Bloch bands, these local properties are encoded in
the quantum geometric tensor. Its real part defines the quantum metric, while
its imaginary part gives the Berry curvature. The Berry curvature determines
topological responses such as the Hall conductivity
\cite{Thouless:1982uq,kohmoto:85}, whereas the quantum metric measures the
distance between neighboring Bloch states in momentum space
\cite{PhysRevB.90.165139,Kolodrubetz:2017aa}.

In recent years, the quantum metric has attracted broad interest in several
contexts. It constrains the spatial localization of Wannier functions and is
closely related to the gauge-invariant part of the Wannier spread
\cite{Marzari:1997aa,PhysRevLett.82.370,PhysRevB.62.1666}. It also appears in
optical and nonlinear responses, where the geometry of Bloch states contributes
to response coefficients independently of the band dispersion
\cite{doi:10.1126/sciadv.1501524,PhysRevB.108.L201405,doi:10.1126/science.adf1506}.
In superconducting systems, the quantum metric gives an important contribution
to the superfluid weight, especially in flat or nearly flat bands
\cite{Peotta:2015aa,PhysRevB.95.024515}. 
\cin{Related higher-order quantum geometric tensors, constructed from higher-order
gauge-invariant cumulants, have also been discussed in Ref.~\cite{PhysRevA.108.032218}.}
These developments show that the
quantum metric is not merely a formal geometric object, but a physical quantity
that controls localization, transport, and response properties of quantum
matter.

Among systems with nontrivial quantum geometry, Chern bands provide a
particularly important class of examples. In a two-dimensional Chern band, the
quantum metric and Berry curvature satisfy a geometric inequality whose
integral gives a lower bound determined by the Chern number. The lowest Landau
level is the canonical example of an ideal Chern band: it saturates this
geometric lower bound and has uniform Berry curvature
\cite{Ozawa:2021vs,PhysRevB.104.045104,PhysRevResearch.6.033238,1zg9-qbd6}. This motivates the search
for lattice Chern bands whose quantum geometry is as close as possible to that
of a Landau level.

Recently, several approaches have been proposed to deform quantum states toward
ideal quantum geometry while preserving their topological properties
\cite{vdp3-dcht,5vhk-7x54,mera2025relaxationidealchernband,
zhao2026idealbandstightbindingmodels}.
In our previous work, we introduced gradient flows generated by geometric
actions constructed from the quantum metric, such as the trace action and the
quantum volume \cite{5vhk-7x54}. These actions decrease monotonically along the
corresponding flow equations and the flow drives the projector toward 
configurations saturating the Bogomolny bound. However, numerical results for
the Wilson-Dirac model revealed that the Berry curvature does not become 
uniform under the metric flow. Instead, it tends to develop a sharply peaked
structure. A similar tendency has also been observed in a different flow method
\cite{mera2025relaxationidealchernband}.

Motivated by this observation, we introduce in this paper a gradient flow
generated by a combined geometric action. The action consists of the
quantum-metric term and the square of the Berry curvature. The metric term
drives the projector toward Bogomolny saturation, whereas the curvature term
suppresses spatial fluctuations of the Berry curvature. We show that these two
requirements compete with each other in finite-dimensional lattice models. As a
result, the gradient flow converges to a nontrivial fixed point that balances
Bogomolny saturation and curvature uniformity.

This competition is closely related to the geometric structure of ideal Chern
bands. It is known that Bogomolny saturation is associated with a K\"ahler, or
holomorphic, structure of the Bloch states, and that Landau levels and their
higher-Chern-number analogs provide canonical realizations of ideal quantum
geometry \cite{Ozawa:2021vs,PhysRevB.104.045104,PhysRevResearch.6.033238,1zg9-qbd6}. 
Building on this
viewpoint, we formulate a simple finite-dimensional obstruction for rank-one
projectors. We show that a finite-dimensional holomorphic Bloch frame cannot
simultaneously realize Bogomolny saturation and exactly uniform Berry
curvature. 
The lowest Landau level avoids this obstruction because its Bloch wave
function is naturally a holomorphic vector taking values in an
infinite-dimensional Hilbert space.

Finally, we apply the combined gradient flow to the Wilson-Dirac model.  We 
show that increasing the weight of the Berry-curvature term makes the Berry
curvature nearly uniform, while increasing the weight of the metric term reduces
the Bogomolny defect.  We further construct a short-range truncated flattened
Hamiltonian from the flowed projector and obtain a lattice model with nearly
flat bands and nearly uniform Berry curvature.  We also examine the Hofstadter
model and demonstrate that the two terms play analogous roles in a Chern band
with a higher Chern number.

The rest of this paper is organized as follows.  Section~\ref{s:action}
introduces the geometric formulation and the action functional.
In Sec.~\ref{s:gradient_flow}, we derive the corresponding
gradient-flow equation.  In Sec.~\ref{s:bogomolny}, we discuss the
Bogomolny bound, its relation to uniform Berry curvature, and the
no-go theorem.  Section~\ref{s:numerics} presents numerical
applications to lattice models.  Section~\ref{s:conclusion} summarizes
the results.

\section{Quantum geometry and gradient-flow action}\label{s:action}

When the Bogomolny bound is saturated, the wave functions acquire a
holomorphic structure. To discuss this property, we first introduce a complex
structure on the Brillouin zone. For this purpose, it is convenient to
distinguish between oblique momentum coordinates associated with reciprocal
lattice vectors and orthonormal coordinates. In this section, we fix our
notation for these coordinates and then define the geometric actions that
generate the gradient flow.

\subsection{Notation for reciprocal space}

We denote by \(\bm b_\mu\) \((\mu=1,2)\) an oblique basis of wave-vector
space, or equivalently a reciprocal-lattice basis, and write
$\bm k=\bm b_\mu k^\mu $.
Since this basis is not necessarily orthonormal, we introduce the dual basis
\(\bm b^\mu\) by
\begin{alignat}1
\bm b^\mu\cdot\bm b_\nu=\delta^\mu_\nu .
\label{DuaBas}
\end{alignat}
In Sec.~\ref{s:LLL}, the dual vectors are used as primitive translation
vectors. The coefficients \(k^\mu\) are called oblique momentum coordinates.

We also introduce orthonormal vectors \(\bm e_a\), satisfying
\(\bm e_a\cdot\bm e_b=\delta_{ab}\), and the dual basis
\(\bm e^a=\delta^{ab}\bm e_b\). The wave vector can then be written in terms of
the orthonormal components as
\begin{alignat}1
\bm k=\bm e_a q^a=\bm b_\mu k^\mu .
\end{alignat}
This gives the relation between the orthonormal coordinates \(q^a\) and the
oblique coordinates \(k^\mu\),
\begin{alignat}1
q^a= e^a{}_\mu k^\mu,
\quad
k^\mu=e_a{}^\mu q^a ,
\label{OrtCor}
\end{alignat}
where \(e^a{}_\mu\) and \(e_a{}^\mu\) are defined by
\begin{alignat}1
e^a{}_\mu=\bm e^a\cdot\bm b_\mu,
\quad
e_a{}^\mu=\bm e_a\cdot\bm b^\mu .
\label{VieBei}
\end{alignat}
These two matrices are inverse to each other. Thus we have 
$\bm b_\mu=e^a{}_\mu\bm e_a$ and
$\bm e_a=e_a{}^\mu\bm b_\mu$.

The metric on the Brillouin zone is induced by
\begin{alignat}1
\bm k^2=\delta_{ab}q^a q^b=
\delta_{ab}e^a{}_\mu e^b{}_\nu k^\mu k^\nu=h_{\mu\nu}k^\mu k^\nu ,
\end{alignat}
where
\begin{alignat}1
h_{\mu\nu}=
\bm b_\mu\cdot\bm b_\nu=\delta_{ab}e^a{}_\mu e^b{}_\nu .
\label{HandVie}
\end{alignat}
The volume element is therefore given by
\begin{alignat}1
d\mu_h=\sqrt{h}\,dk^1\wedge dk^2,
\end{alignat}
where \(h=\det h_{\mu\nu}\).

\subsection{Quantum geometric tensor}

Consider the wave function \(\psi_{i,n}(k)\), where
\(i=1,\dots,N\) labels the orbital in the unit cell and \(n\) labels the energy
eigenstate. We introduce the occupied \(r\)-multiplet
$\Psi_i(k)=\left(\psi_{i,1}(k),\cdots,\psi_{i,r}(k)\right)$.
The projector onto the occupied subspace is then given by
$P_{ij}(k)=\sum_{n=1}^r\psi_{i,n}(k)\psi_{j,n}^*(k)
=\Psi_i(k)\Psi_j^\dagger(k)$,
which will often be denoted simply by \(P(k)\). Note that \(P\) obeys
\begin{alignat}1
P^2=P,\quad P^\dagger=P .
\end{alignat}

The quantum geometric tensor is defined by
\begin{alignat}1
Q_{\mu\nu}(k)=\Tr P\partial_\mu P\partial_\nu P ,
\label{QuaGeoTen}
\end{alignat}
where \(\Tr\) denotes the trace over the orbital space,
\(\Tr A=\sum_{i=1}^N A_{ii}\), and
\(\partial_\mu\equiv \partial/\partial k^\mu\) denotes the derivative with
respect to the momentum coordinate.

The quantum metric and the Berry curvature are defined by the symmetric and
antisymmetric parts of the quantum geometric tensor (\ref{QuaGeoTen}):
\begin{alignat}1
Q_{\mu\nu}&=\frac{1}{2}\Tr P\{\partial_\mu P,\partial_\nu P\}
+\frac{1}{2}\Tr P[\partial_\mu P,\partial_\nu P]
\nonumber\\
&=g_{\mu\nu}+\frac{1}{2}F_{\mu\nu},
\end{alignat}
where
\begin{alignat}1
g_{\mu\nu}(k)&=
\frac{1}{2}\Tr \partial_\mu P\partial_\nu P,
\nonumber\\
F_{\mu\nu}(k)&=\Tr P[\partial_\mu P,\partial_\nu P].
\label{DefGandF}
\end{alignat}
Note that \(g_{\mu\nu}\) is real, whereas \(F_{\mu\nu}\) is purely imaginary.

We define the Berry-curvature two-form by
\begin{alignat}1
{\cal F}=\frac{1}{2}F_{\mu\nu}dk^\mu\wedge dk^\nu=-i\Omega_h\, d\mu_h .
\label{Cur2For}
\end{alignat}
With this convention, the Chern number is
\begin{alignat}1
C=\frac{i}{2\pi}\int_{T^2} {\cal F}=\frac{1}{2\pi}\int_{T^2} \Omega_h\,d\mu_h ,
\end{alignat}
where \(T^2\) denotes the Brillouin-zone torus spanned by \(k^\mu\).

When the dependence on the Brillouin-zone metric should be emphasized, we write
the Berry curvature and volume form as \(\Omega_h\) and \(d\mu_h\). In the
following, we mostly suppress the subscript and write
\(\Omega\equiv\Omega_h\) and \(d\mu\equiv d\mu_h\).

\subsection{Action driving the gradient flow}

In Ref.~\cite{5vhk-7x54}, we studied gradient flows generated by geometric
actions constructed from the quantum metric, such as the trace action and the
quantum volume. In the present paper, we focus on the trace action and add a
Berry-curvature term, which tends to suppress fluctuations of \(\Omega\) over
the Brillouin zone. We define the action as
\begin{alignat}1
S=\alpha S_g+\beta S_\Omega=
\frac{1}{2}\int_{T^2}\left[\alpha\tr g(k)+\beta\Omega^2(k)\right]d\mu ,
\label{DelSg}
\end{alignat}
where \(\alpha\) and \(\beta\) are nonnegative parameters. Here \(\tr g\)
denotes the trace with respect to the Brillouin-zone metric,
$ \tr g=g^\mu{}_\mu=h^{\mu\nu}g_{\mu\nu}$.

\section{Gradient flow equation}\label{s:gradient_flow}

The flow parameter introduced below is an auxiliary parameter and should not be regarded as physical time. The gradient flow is constructed so that the action decreases monotonically while the projector condition \(P^2=P\) is preserved.

\subsection{First variation of the action}

In this subsection, we derive the first variation of the action separately for \(S_g\) and
\(S_\Omega\).

The variation of \(S_g\) was already derived in
Ref.~\cite{5vhk-7x54}. It is given by
\begin{alignat}1
\delta S_g=\int \Tr K_g\delta P \,d\mu,
\end{alignat}
where \(K_g\) is the Hermitian matrix
\begin{alignat}1
K_g=-\frac{1}{2}\Delta P=-\frac{1}{2}\partial^\mu\partial_\mu P .
\end{alignat}

We next derive the variation of \(S_\Omega\). We introduce the Levi-Civita
tensor by
\begin{alignat}1
\epsilon^{\mu\nu}=\frac{\varepsilon^{\mu\nu}}{\sqrt h},
\quad
(\varepsilon^{12}=-\varepsilon^{21}=1) .
\end{alignat}
Then the Berry curvature is written as
\begin{alignat}1
\Omega=i\epsilon^{\mu\nu}\Tr P\partial_\mu P\partial_\nu P .
\end{alignat}
The variation of \(\Omega\) is
\begin{alignat}1
\delta\Omega&=\delta
\left(i\epsilon^{\mu\nu}\Tr P\partial_\mu P\partial_\nu P\right)
\nonumber\\
&=i\epsilon^{\mu\nu}\Tr
\left(
\delta P\partial_\mu P\partial_\nu P
+P\partial_\mu\delta P\partial_\nu P
+P\partial_\mu P\partial_\nu \delta P
\right).
\end{alignat}
After integrating by parts on the Brillouin-zone torus, we obtain
\begin{alignat}1
\delta S_\Omega
&=\int \Omega\delta\Omega \,d\mu
\nonumber\\
&=i\epsilon^{\mu\nu}\int
\Tr\big[
\Omega\partial_\mu P \partial_\nu P
-\partial_\mu (\Omega \partial_\nu P P)
\nonumber\\
&\qquad\qquad-\partial_\nu( \Omega P\partial_\mu P)
\big]\delta P \,d\mu
\nonumber\\
&=\int\Tr K_\Omega\delta P\,d\mu,
\label{DelSOme}
\end{alignat}
where \(K_\Omega\) is the Hermitian matrix
\begin{alignat}1
K_\Omega
\equiv i\epsilon^{\mu\nu}
\left(3\Omega\partial_\mu P \partial_\nu P+\partial_\mu \Omega [P,\partial_\nu P]\right).
\end{alignat}
The first term in \(K_\Omega\) is block diagonal with respect to \(P\) and 
therefore does not contribute to the projected gradient, as shown in
Sec.~\ref{s:GraFloEqu}.

Combining Eqs.~\eqref{DelSg} and \eqref{DelSOme}, we obtain
\begin{alignat}1
\delta S=\int\Tr K\delta P\,d\mu,
\quad
K=\alpha K_g+\beta K_\Omega .
\label{DelS}
\end{alignat}

\subsection{Gradient flow equation}\label{s:GraFloEqu}

Let us introduce an auxiliary parameter \(t\), which describes a one-parameter
deformation of the projector \(P(t,k)\). A naive gradient-flow equation would be
\begin{alignat}1
\dot P=-\frac{\delta S}{\delta P}.
\end{alignat}
However, this equation does not take into account the constraint \(P^2=P\).
Differentiating \(P^2=P\) with respect to \(t\), we find that \(\dot P\) is
off-diagonal with respect to \(P\):
$P\dot P P=0,\,(1-P)\dot P(1-P)=0.$
Therefore, the right-hand side of the naive flow equation, \(-K\), should be
projected onto the off-diagonal subspace. The corresponding projection is
$\Pi_P K\equiv PK(1-P)+(1-P)KP= [P,[P,K]]$.
Thus, the gradient flow equation on the projector manifold is given by
$\dot P=-\Pi_P K$ 
or, equivalently, 
\begin{alignat}1
\dot P=-[P,J], \quad J=[P,K].
\label{GraFloEqu}
\end{alignat}
Since \(K^\dagger=K\), the generator \(J\) is anti-Hermitian,
\(J^\dagger=-J\).

Let us compute \(J\) explicitly. Using the expression for \(K\) in Eq. \eqref{DelS}, we obtain
\begin{alignat}1
J&=[P,K]=-\frac{\alpha}{2}[P,\Delta P]+\beta i\epsilon^{\mu\nu}\partial_\mu\Omega
[P,[P,\partial_\nu P]]
\nonumber\\
&=-\frac{\alpha}{2}[P,\Delta P]+\beta i\epsilon^{\mu\nu}\partial_\mu\Omega \partial_\nu P .
\label{J}
\end{alignat}
Here we have used
\([P,\partial_\mu P\partial_\nu P]=0\), since
\(\partial_\mu P\partial_\nu P\) is block diagonal with respect to \(P\), and
\([P,[P,\partial_\nu P]]=\partial_\nu P\).

The projected flow still decreases the action monotonically. It follows from
Eqs.~\eqref{DelS} and \eqref{GraFloEqu} that
\begin{alignat}1
\dot S&=\int \Tr K\dot P\,d\mu=-\int\Tr K[P,J]\,d\mu
\nonumber\\
&=-\int\Tr [K,P]J\,d\mu =-\int\Tr J^\dagger J\,d\mu \le0.
\end{alignat}

\subsection{Berry curvature and quantum metric under gradient flow}

So far we have derived the gradient-flow equation \eqref{GraFloEqu}, which
determines the time evolution of the projector \(P(t,k)\). Once \(P(t,k)\) is
known, the quantum metric \(g_{\mu\nu}\) and the Berry curvature \(\Omega\)
can be computed at each flow time. Nevertheless, it is useful to derive the
induced flow equations for these geometric quantities directly, because they
clarify the roles played by \(S_g\) and \(S_\Omega\) in the gradient flow. 

\subsubsection{Gradient flow equation for the Berry curvature}

We first derive the flow equation for the Berry curvature. Substituting
Eq.~\eqref{GraFloEqu} into \(\dot\Omega\), we obtain
\begin{widetext}
\begin{alignat}1
\dot\Omega
&=i\epsilon^{\mu\nu}\Tr
\left(\dot P\partial_\mu P\partial_\nu P
+P\partial_\mu \dot P\partial_\nu P
+P\partial_\mu P\partial_\nu \dot P
\right)
\nonumber\\
&=
-i\epsilon^{\mu\nu}\Tr
\left(
[P,J]\partial_\mu P\partial_\nu P
+P\partial_\mu [P,J]\partial_\nu P
+P\partial_\mu P\partial_\nu [P,J]
\right).
\end{alignat}
Using the trace identity \(\Tr[A,B]C=\Tr A[B,C]\), this becomes
\begin{alignat}1
\dot\Omega&=
-i\epsilon^{\mu\nu}\Tr
\Big(
[\partial_\mu P\partial_\nu P,P]J
+\left(
[\partial_\nu P P,\partial_\mu P]
+[P\partial_\mu P,\partial_\nu P]
\right)J
+[\partial_\nu P P,P]\partial_\mu J
+[P\partial_\mu P,P]\partial_\nu J
\Big).
\end{alignat}
\end{widetext}
Further using $[\partial_\nu P P,\partial_\mu P]=[\partial_\nu P,P\partial_\mu P]$, we find
\begin{alignat}1
\dot\Omega
&=-i\epsilon^{\mu\nu}\Tr\left(\partial_\nu P P\partial_\mu J
-P\partial_\mu P\partial_\nu J\right)
\nonumber\\
&=-i\epsilon^{\mu\nu}\Tr\partial_\nu P\partial_\mu J=-\partial_\mu j^\mu,
\label{FloOme}
\end{alignat}
where
\begin{alignat}1
j^\mu=i\epsilon^{\mu\nu}\Tr\partial_\nu P J .
\label{OmeCur}
\end{alignat}
Thus the Berry curvature obeys a continuity equation. Since the Brillouin zone
is a torus, the integral of the right-hand side of Eq. \eqref{FloOme} vanishes. Therefore, the Chern
number, namely the integral of \(\Omega\) over the Brillouin zone, is conserved
along the gradient flow.

The current \(j^\mu\) consists of two contributions associated with \(S_g\) and
\(S_\Omega\). We compute these contributions separately using Eq.~\eqref{J}.
The contribution from the metric action \(S_g\) is
\begin{alignat}1
\frac{2}{\alpha}j^\mu_g&=
-i\epsilon^{\mu\nu}\Tr\partial_\nu P [P,\Delta P]=-i\epsilon^{\mu\nu}\Tr P[\Delta P,\partial_\nu P]
\nonumber\\
&=i\epsilon^{\mu\nu}\left(Q_{\nu,\rho}{}^\rho-Q_{\rho}{}^\rho{}_{,\nu}\right),
\end{alignat}
where we have introduced \cite{qscv-qxqt}
\begin{alignat}1
Q_{\mu,\nu\rho}
&\equiv\Tr P\partial_\mu P\partial_\nu\partial_\rho P,
\nonumber\\
Q_{\mu\nu,\rho}
&\equiv \Tr P\partial_\mu\partial_\nu P\partial_\rho P=Q_{\rho,\mu\nu}^* .
\end{alignat}
It follows that
\begin{alignat}1
\frac{2}{\alpha}j^\mu_g&=
i\epsilon^{\mu\nu}\left(Q_{\nu,\rho}{}^{\rho}-Q^*_{\nu,\rho}{}^{\rho}\right)=
-2\epsilon^{\mu\nu}\Im Q_{\nu,\rho}{}^\rho .
\end{alignat}
On the other hand, the contribution from \(S_\Omega\) is
\begin{alignat}1
\frac{j^\mu_\Omega}{\beta}
&=
i\epsilon^{\mu\nu}
\Tr\partial_\nu P
\left(
i\epsilon^{\rho\sigma}
\partial_\rho\Omega\partial_\sigma P
\right)
\nonumber\\
&=
-\epsilon^{\mu\nu}\epsilon^{\rho\sigma}
\partial_\rho\Omega
\Tr\partial_\nu P\partial_\sigma P
\nonumber\\
&=
-D^{\mu\nu}\partial_\nu \Omega,
\end{alignat}
where
\begin{alignat}1
D^{\mu\nu}\equiv 2\epsilon^{\mu\rho}\epsilon^{\nu\sigma}g_{\rho\sigma}.
\end{alignat}
In matrix form, this tensor is written as
\begin{alignat}1
D
=\frac{2}{h}
\begin{pmatrix}
g_{22} & -g_{12}\\ -g_{12} & g_{11}
\end{pmatrix}.
\end{alignat}

Combining the two contributions, we obtain the flow equation for the Berry
curvature,
\begin{alignat}1
\dot \Omega=\partial_\mu
\left(\alpha\epsilon^{\mu\nu}\Im Q_{\nu,\rho}{}^\rho+\beta D^{\mu\nu}\partial_\nu\Omega\right).
\label{OmegaFlowFinal}
\end{alignat}
The second term has the form of an anisotropic diffusion term for the Berry
curvature, where \(D^{\mu\nu}\) plays the role of a diffusion tensor determined
by the quantum metric. This shows explicitly that the \(S_\Omega\) part of the
action tends to suppress spatial fluctuations of the Berry curvature and drives
\(\Omega\) toward a more uniform distribution.

\subsubsection{Gradient flow equation for the quantum metric}

We next derive the flow equation for the quantum metric. From the definition of
\(g_{\mu\nu}\), we have
\begin{alignat}1
2\dot g_{\mu\nu}=&
\Tr\left(\partial_\mu\dot P\partial_\nu P+\partial_\mu P\partial_\nu\dot P\right)
\nonumber\\
=&-\Tr\left(\partial_\mu [P,J]\partial_\nu P+\partial_\mu P\partial_\nu [P,J]\right)
\nonumber\\
=&-\Tr
\big([\partial_\mu P,J]\partial_\nu P+[P,\partial_\mu J]\partial_\nu P
\nonumber\\
&+\partial_\mu P[\partial_\nu P,J]+\partial_\mu P[P,\partial_\nu J]\big).
\end{alignat}
The first and third terms cancel each other. Therefore,
\begin{alignat}1
2\dot g_{\mu\nu}&=-\Tr\left([\partial_\nu P,P]\partial_\mu J+[\partial_\mu P,P]\partial_\nu J\right)
\nonumber\\
&=\Tr P\partial_\nu P\partial_\mu J-\Tr \partial_\nu P P\partial_\mu J+(\mu\leftrightarrow\nu)
\nonumber\\
&=\Tr P\partial_\nu P\partial_\mu J-\left(\Tr \partial_\mu J^\dagger P\partial_\nu P\right)^*
+(\mu\leftrightarrow\nu).
\end{alignat}
Using the anti-Hermiticity of \(J\), we obtain
\begin{alignat}1
2\dot g_{\mu\nu}&=\Tr P\partial_\nu P\partial_\mu J+
\left(\Tr \partial_\mu J P\partial_\nu P\right)^*+(\mu\leftrightarrow\nu)
\nonumber\\
&=2\Re\left(\Tr P\partial_\nu P\partial_\mu J\right)+(\mu\leftrightarrow\nu).
\label{MetricFlowGeneral}
\end{alignat}
This is the metric counterpart of Eq.~\eqref{FloOme}.

Substituting Eq.~\eqref{J} into Eq.~\eqref{MetricFlowGeneral}, we obtain
\begin{alignat}1
2\dot g_{\mu\nu}
=&-\alpha\Re\Tr P\partial_\nu P\partial_\mu([P,\Delta P])
\nonumber\\
&-2\beta\epsilon^{\rho\sigma}\Im\Tr P\partial_\nu P\partial_\mu
\left(\partial_\rho\Omega\partial_\sigma P\right)
+(\mu\leftrightarrow \nu).
\end{alignat}
The term proportional to \(\beta\) can be rewritten as
\begin{alignat}1
&-2\beta \epsilon^{\rho\sigma}
\left(\partial_\mu\partial_\rho\Omega\Im \Tr P\partial_\nu P\partial_\sigma P
+\partial_\rho\Omega\Im \Tr P\partial_\nu P\partial_\mu \partial_\sigma P\right)
\nonumber\\
&=-2\beta \epsilon^{\rho\sigma}
\left(\partial_\mu\partial_\rho\Omega\Im Q_{\nu\sigma}+\partial_\rho\Omega\Im Q_{\nu,\mu\sigma}\right)
\nonumber\\
&=\beta\left(\Omega\partial_\mu\partial_\nu\Omega-
2\epsilon^{\rho\sigma}\partial_\rho\Omega\Im Q_{\nu,\mu\sigma}\right),
\end{alignat}
where \(Q_{\mu\nu}\) in the middle line denotes the quantum geometric tensor defined in
Eq.~\eqref{QuaGeoTen}. Thus the flow equation for the quantum metric is 
\begin{alignat}1
\dot g_{\mu\nu}
=&
-\frac{\alpha}{2}
\Re\Tr
P\partial_\nu P
\partial_\mu([P,\Delta P])
\nonumber\\
&+
\frac{\beta}{2}
\left(
\Omega\partial_\mu\partial_\nu\Omega
-
2\epsilon^{\rho\sigma}
\partial_\rho\Omega
\Im Q_{\nu,\mu\sigma}
\right)
+
(\mu\leftrightarrow \nu).
\label{MetricFlowFinal}
\end{alignat}

Compared with the flow equation for \(\Omega\), the equation for
\(g_{\mu\nu}\) is less transparent. In particular, the term proportional to
\(\beta\) is not a diffusion term for the quantum metric. Rather, it describes
how the metric is deformed as a consequence of the curvature-flattening flow.
Therefore, the \(S_\Omega\) part of the flow does not generally preserve the
Bogomolny condition. In finite-dimensional lattice models, this is precisely
where the competition between Bogomolny saturation and curvature uniformity
appears.

Although the total action decreases monotonically, its two components do not
necessarily decrease separately. This can be seen explicitly as follows. Let
\(J_g=[P,K_g]\) and \(J_\Omega=[P,K_\Omega]\) be the projected gradients
generated by \(S_g\) and \(S_\Omega\), respectively. Since
$J=\alpha J_g+\beta J_\Omega$,
the metric action changes along the combined flow as
\begin{alignat}1
\dot S_g&=-\int\Tr J_g^\dagger J\,d\mu
\nonumber\\
&=-\alpha
\int\Tr J_g^\dagger J_g\,d\mu-
\beta\int\Tr J_g^\dagger J_\Omega\,d\mu.
\label{SgChange}
\end{alignat}
The first term is nonpositive, but the second term has no definite sign.
Therefore, the curvature-flattening part of the flow can increase the metric
action, or equivalently the Bogomolny defect, even though the total action
\(S=\alpha S_g+\beta S_\Omega\) decreases monotonically. This explains why the
flow with larger \(\beta\) can make the Berry curvature more uniform while
remaining farther from Bogomolny saturation.

\section{Lower bound of the action}\label{s:bogomolny}

In this section, we derive lower bounds for the two terms in the action. These
bounds clarify the ideal geometry toward which the gradient flow drives the
projector \(P\). The lower bound of the metric action is the analog of the 
Bogomolny bound for maps into the Grassmannian. In the context of Bloch bands,
this bound is closely related to ideal Chern-band geometry, the K\"ahler
condition, and the existence of a holomorphic Bloch frame
\cite{Ozawa:2021vs,PhysRevB.104.045104,PhysRevResearch.6.033238,1zg9-qbd6,5vhk-7x54}.

\subsection{Bogomolny bound for \(S_g\)}

It is convenient to derive the lower bound of \(S_g\) using complex
coordinates. According to Eq. \eqref{OrtCor}, we first introduce orthonormal coordinates \(q^a\) by
\begin{alignat}1
q^a=e^a{}_\mu k^\mu,
\quad
\frac{\partial}{\partial q^a}=e_a{}^\mu
\frac{\partial}{\partial k^\mu}.
\end{alignat}
We then define
\begin{alignat}1
q=q^1+iq^2,\quad
\bar q=q^1-iq^2,
\end{alignat}
and
\begin{alignat}1
\partial\equiv\frac{\partial}{\partial q}=\frac{1}{2}(\partial_1-i\partial_2),
\quad
\bar\partial\equiv\frac{\partial}{\partial \bar q}=\frac{1}{2}(\partial_1+i\partial_2).
\end{alignat}
Note that, in this orthonormal space, $\sqrt{h}=1$. 
Using the identities
\begin{alignat}1
\partial^a P\partial_a P=
\left\{
\begin{array}{l}
4\,\partial P\bar\partial P-i\epsilon^{ab}\partial_aP\partial_bP
\\
4\,\bar\partial P\partial P+i\epsilon^{ab}\partial_aP\partial_bP
\end{array}
\right.,
\end{alignat}
and 
\begin{alignat}1
\Tr \partial^a P\partial_aP=2\Tr P\partial^aP\partial_aP,
\end{alignat}
we can rewrite \(S_g\) as
\begin{alignat}1
S_g&=
\frac{1}{4}\int\Tr \partial^a P\partial_a P\,d\mu=\frac{1}{2}\int\Tr P\partial^a P\partial_a P\,d\mu
\nonumber\\
&=2\int\Tr\left\{
\begin{array}{l}
P\partial P\bar\partial P
\\
P\bar\partial P\partial P
\end{array}
\right\}d\mu
\nonumber\\
&\qquad\mp\frac{i}{2}\int\epsilon^{ab}\Tr P\partial_aP\partial_bP\,dq^1\wedge dq^2.
\end{alignat}
The first term above is nonnegative, since 
\begin{alignat}1
\Tr
\left\{
\begin{array}{l}
P\partial P\bar\partial P
\\
P\bar\partial P\partial P
\end{array}
\right\}
=
\left\{
\begin{array}{l}
\norm{\bar\partial PP}^2
\\
\norm{\partial PP}^2
\end{array}
\right\},
\end{alignat}
where \(\norm{A}^2\equiv \Tr A^\dagger A\) denotes the squared  norm of  $A$.
The second term gives the Chern number 
\begin{alignat}1
\cin{\frac{i}{2}}\int\epsilon^{ab}\Tr P\partial_aP\partial_bP\,dq^1\wedge dq^2
&=
\cin{\frac{i}{2}}\int\Tr P\partial_aP\partial_bP\,dq^a\wedge dq^b
\nonumber\\
&=
\cin{\frac{i}{2}}\int\Tr P\partial_\mu P\partial_\nu P\,dk^\mu\wedge dk^\nu
\nonumber\\
&=\pi C.
\end{alignat}
Therefore, the Bogomolny bound for \(S_g\) is \cite{PhysRevB.90.165139,Ozawa:2021vs}
\begin{alignat}1
S_g\ge\pi |C|.
\end{alignat}
The bound is saturated if and only if
\begin{alignat}1
\left\{
\begin{array}{ll}\bar\partial PP=0,&(C<0),\\
\partial PP=0,&(C>0).
\end{array}
\right.
\label{BPScon}
\end{alignat}
This is the local ideal quantum geometry condition, often formulated as the
K\"ahler or holomorphicity condition in the literature on ideal Chern bands
\cite{Ozawa:2021vs,PhysRevB.104.045104,PhysRevResearch.6.033238,1zg9-qbd6}.

\subsubsection{Bogomolny-saturating solutions}

Let us define
$X_a=\partial_a P P$.
In terms of \(X_a\), the quantum metric and the Berry curvature in the orthonormal frame are written as
\begin{alignat}1
g_{ab}&=\frac{1}{2}
\left(\Tr X_a^\dagger X_b+\Tr X_b^\dagger X_a\right),
\nonumber\\
\Omega&=i\epsilon^{ab}\Tr P\partial_aP\partial_bP
=i\Tr\left(X_1^\dagger X_2-X_2^\dagger X_1\right).
\label{CurOrt}
\end{alignat}
On the other hand, the Bogomolny condition \eqref{BPScon} is equivalent to
\begin{alignat}1
X_1\pm iX_2=0 .
\end{alignat}
Substituting this condition into Eq.~\eqref{CurOrt}, we obtain
\begin{alignat}1
g_{11}&=g_{22}=\Tr X_1^\dagger X_1,
\quad
g_{12}=g_{21}=0,
\nonumber\\
\Omega&=\mp 2\Tr X_1^\dagger X_1 .
\end{alignat}
Therefore, when the Bogomolny bound is saturated, the quantum metric satisfies
\begin{alignat}1
g_{ab}=\frac{|\Omega|}{2}\delta_{ab}
\end{alignat}
in the orthonormal frame. In the original \(k^\mu\) coordinates, this relation
becomes
\begin{alignat}1
g_{\mu\nu}=\frac{|\Omega|}{2}h_{\mu\nu},
\end{alignat}
where we have used
$g_{\mu\nu}=e^a{}_\mu e^b{}_\nu g_{ab}$.

We next solve the holomorphic or antiholomorphic condition \eqref{BPScon}. 
Let \(W\) be an \(N\times r\) matrix whose columns are unnormalized occupied
states, where \(r\) is the number of occupied bands. The projector is written as \cite{PhysRevB.104.045104}
\begin{alignat}1
P=WH^{-1}W^\dagger,
\quad
H=W^\dagger W.
\label{BPSsol}
\end{alignat}
It follows that \(PW=W\) and \(W^\dagger P=W^\dagger\). Differentiating these
relations, we obtain
\begin{alignat}1
(1-P)dW=dP\,W,
\quad
dW^\dagger(1-P)=W^\dagger dP .
\label{PWrel}
\end{alignat}
Combining these relations with the Bogomolny condition \eqref{BPScon}, we find
\begin{alignat}1
(1-P)
\left\{
\begin{array}{l}\bar\partial W\\\partial W\end{array}
\right.=0
\quad
\begin{array}{l}(C<0)\\(C>0)\end{array}.
\label{HolW_1}
\end{alignat}
This implies that \(\bar\partial W\) or \(\partial W\) lies in the image of
\(W\). Namely,
$\bar\partial W=WA_{\bar q} \,(C<0)$ or
$\partial W=WA_q\,(C>0)$,
where \(A_{\bar q}\) and \(A_q\) are \(r\times r\) matrices. Locally, by an
appropriate 
gauge transformation of \(W\), we can choose
\begin{alignat}1
\left\{
\begin{array}{l}
\bar\partial W\\ \partial W \end{array}
\right.=0
\quad
\begin{array}{l} (C<0) \\ (C>0) \end{array}.
\label{HolW_2}
\end{alignat}
Thus Bogomolny saturation is equivalent, locally, to the existence of a 
holomorphic frame for \(C<0\) or an antiholomorphic frame for \(C>0\). This 
viewpoint is standard in the geometric formulation of ideal Chern bands
\cite{Ozawa:2021vs,PhysRevB.104.045104,PhysRevResearch.6.033238,1zg9-qbd6}.

For a projector of the form \eqref{BPSsol} satisfying \eqref{HolW_2}, the Berry
curvature is determined by the Hermitian matrix \(H=W^\dagger W\) \cite{PhysRevB.104.045104}.
To see this,
we introduce the non-Abelian Berry connection
\begin{alignat}1
{\cal A}=H^{-1}W^\dagger dW .
\label{BerCon}
\end{alignat}
The corresponding non-Abelian curvature two-form is
\begin{alignat}1
\tilde{\cal F}&=d{\cal A}+{\cal A}\wedge{\cal A}
\nonumber\\
&=H^{-1}dW^\dagger(1-P)\wedge dW
\nonumber\\
&=H^{-1}W^\dagger dP\wedge dP\,W,
\end{alignat}
where we have used Eq.~\eqref{PWrel} in the last line. Taking the trace over
the occupied-band indices, we obtain the Abelian Berry curvature two-form
\begin{alignat}1
{\cal F}\equiv\cin{\Trr}\tilde{\cal F}=\Tr P\,dP\wedge dP ,
\end{alignat}
\cin{where $\Trr$ denotes the trace over the energy band indices.}

For the holomorphic or antiholomorphic frame \eqref{HolW_2}, the connection 
\eqref{BerCon} becomes
\begin{alignat}1
{\cal A}=
\left\{
\begin{array}{ll}
H^{-1}\partial H\,dq \quad& (C<0)
\\
H^{-1}\bar\partial H\,d\bar q &(C>0)
\end{array} .
\right.
\end{alignat}
Therefore,
\begin{alignat}1
\tilde{\cal F}=
\left\{
\begin{array}{ll}
-\bar\partial(H^{-1}\partial H)\,dq\wedge d\bar q \quad& (C<0)
\\
+\partial(H^{-1}\bar\partial H)\,dq\wedge d\bar q &(C>0)
\end{array}
\right. .
\end{alignat}
Using the identity
$\delta\ln\Detr M=\Trr\left(M^{-1}\delta M\right)$, where $\cin{\Detr}$ denotes the determinant associated 
with $\cin{\Trr}$, 
we find the Abelian curvature
\begin{alignat}1
{\cal F}=
\left\{
\begin{array}{ll}
-\bar\partial\partial\ln\cin{\Detr} H\,dq\wedge d\bar q \quad &(C<0)
\\
+\bar\partial\partial\ln\cin{\Detr} H\,dq\wedge d\bar q&(C>0)
\end{array}
\right. .
\end{alignat}
Using Eq. (\ref{Cur2For}) as well as 
$d\mu=dq^1\wedge dq^2=\frac{i}{2}dq\wedge d\bar q$,
we obtain
\begin{alignat}1
\Omega=
\left\{
\begin{array}{ll}
-2\bar\partial\partial\ln\cin{\Detr} H \quad& (C<0)
\\
+2\bar\partial\partial\ln\cin{\Detr} H &(C>0)
\end{array}
\right. .
\label{OmegaHdetH}
\end{alignat}

\subsection{Bound for \(S_\Omega\)}

The curvature part of the action also has a simple lower bound. From the
Cauchy-Schwarz inequality, 
\begin{alignat}1
\left(\int_{T^2}\Omega_h^2\,d\mu_h\right)\left(
\int_{T^2}1\,d\mu_h\right)
\ge\left(\int_{T^2}\Omega_h\,d\mu_h\right)^2=(2\pi C)^2 .
\end{alignat}
Since the volume of the Brillouin zone is 
$\int_{T^2}\,d\mu_h=(2\pi)^2\sqrt h$,
we obtain
\begin{alignat}1
S_\Omega=
\frac{1}{2}\int_{T^2}\Omega_h^2\,d\mu_h \ge
\frac{1}{2}\frac{(2\pi C)^2}{(2\pi)^2\sqrt h}=\frac{C^2}{2\sqrt h}.
\end{alignat}
The equality is saturated when \(\Omega_h\) is constant:
\begin{alignat}1
\bar\Omega_h=\frac{C}{2\pi\sqrt h}.
\label{MeaOme}
\end{alignat}
Here, we write the subscript \(h\) explicitly in \(\Omega_h\) to emphasize
that the Berry-curvature density depends on the Brillouin-zone metric
through its definition with respect to the measure \(d\mu_h\) in
Eq.~\eqref{Cur2For}.

\subsection{Bound for the total action}

Combining the two inequalities, we obtain
\begin{alignat}1
S=\alpha S_g+\beta S_\Omega \ge \alpha\pi |C| + \beta\frac{C^2}{2\sqrt h}.
\label{TotS}
\end{alignat}
This is a formal lower bound obtained by adding the separate lower bounds for
\(S_g\) and \(S_\Omega\). Its saturation requires that both bounds be saturated
simultaneously. Therefore, for a projector parametrized as in Eq.~\eqref{BPSsol},
one needs an appropriate holomorphic or antiholomorphic frame satisfying 
Eq.~\eqref{HolW_2}, together with the uniform-curvature condition. Using
Eq.~\eqref{OmegaHdetH}, this condition is written as
\begin{alignat}1
\bar\partial\partial\ln\cin{\Detr} H=\frac{|C|}{4\pi\sqrt h}.
\label{HolFla}
\end{alignat}

\subsubsection{No-go theorem for simultaneous saturation in rank-one bands}
\label{s:nogo}

We now show that the formal lower bound in Eq.~\eqref{TotS} cannot be
saturated, in general, by finite-dimensional rank-one projectors. The following
argument is closely related to known obstructions to exactly ideal Chern bands
in finite-band lattice models
\cite{Ozawa:2021vs,PhysRevB.104.045104,PhysRevResearch.6.033238,1zg9-qbd6,zhao2026idealbandstightbindingmodels}.
Here we present a simple local version adapted to the present formulation.

Suppose that both the Bogomolny condition \eqref{HolW_2} and the uniform-curvature condition \eqref{HolFla}
are satisfied. On a simply connected coordinate patch, the general solution of Eq. \eqref{HolFla} is
\begin{alignat}1
\ln\cin{\Detr} H=\frac{|C|}{4\pi\sqrt h}q\bar q+\phi(q)+\overline{\phi(q)},
\end{alignat}
where \(\phi(q)\) is a holomorphic function. Hence
\begin{alignat}1
\cin{\Detr} H=\cin{\Detr} W^\dagger W=
e^{\frac{|C|}{4\pi\sqrt h}q\bar q}
|f(q)|^2,
\end{alignat}
where $f(q)=e^{\phi(q)}$.
For a holomorphic frame \(W=W(q)\), or an antiholomorphic frame 
\(W=W(\bar q)\), the holomorphic factor can be removed by a 
gauge transformation,
$W(q)\longrightarrow W(q)e^{-\phi(q)/r}$ or
$W(\bar q) \longrightarrow W(\bar q)e^{-\overline{\phi(q)}/r}$.
Thus the simultaneous saturation condition reduces to
\begin{alignat}1
\cin{\Detr} W^\dagger W=e^{\frac{|C|}{4\pi\sqrt h}q\bar q}.
\label{UniCur}
\end{alignat}

We now show that this condition cannot be satisfied by a finite-dimensional
rank-one holomorphic frame. Let \(r=1\). Then \(W\) is an \(N\)-component
holomorphic vector,
$W^T=(w_1,w_2,\dots,w_N)$.
Locally, each component has a Taylor expansion,
\begin{alignat}1
w_i(q)=\sum_{n=0}^{\infty}\frac{w_i^{(n)}}{n!}q^n,
\end{alignat}
where $w_i^{(n)}= \partial^n w_i/\partial q^n \big|_{q=0} $.
Introducing \(N\)-component vectors
\begin{alignat}1
{\bm w}_n^T
=
(w_1^{(n)},w_2^{(n)},\dots,w_N^{(n)}),
\end{alignat}
the condition \eqref{UniCur} becomes
\begin{alignat}1
\sum_{n,m=0}^{\infty}
\frac{{\bm w}_n^\dagger{\bm w}_m}{n!m!}
\bar q^n q^m=\sum_{n=0}^{\infty}\frac{1}{n!}
\left(\frac{|C|}{4\pi\sqrt h}\right)^n \bar q^n q^n .
\end{alignat}
Comparing the coefficients of \(\bar q^n q^m\), we obtain
\begin{alignat}1
{\bm w}_n^\dagger{\bm w}_m=n! \left(\frac{|C|}{4\pi\sqrt h}\right)^n
\delta_{nm}.
\label{OrthVec}
\end{alignat}
Therefore, for \(C\neq0\), the infinitely many nonzero vectors
\(\{{\bm w}_0,{\bm w}_1,{\bm w}_2,\dots\}\) must be mutually orthogonal. This
is impossible in a finite $N$-dimensional vector space. Hence a
finite-dimensional rank-one holomorphic frame cannot satisfy the
uniform-curvature condition \eqref{UniCur}.

The same argument applies to the anti-holomorphic case. We therefore conclude
that, for finite-dimensional rank-one Chern bands, Bogomolny saturation and
exactly uniform Berry curvature cannot be realized simultaneously. In other
words, the formal lower bound in Eq.~\eqref{TotS} is not attainable in such
models. The lowest Landau level avoids this obstruction because its holomorphic
Bloch frame is naturally infinite dimensional, as will be discussed below. 

\subsubsection{Fixed points of the flow equation}

Let us next discuss the fixed points of the gradient-flow equation
\eqref{GraFloEqu}. A fixed point satisfies
$\dot P=-[P,J]=0$.
Since \(J=[P,K]\) is off diagonal with respect to \(P\), the condition \([P,J]=0\) is equivalent to
$J=0 $.
Using Eq.~\eqref{J}, the fixed-point equation is therefore
\begin{alignat}1
J=-\frac{\alpha}{2}[P,\Delta P]+\beta i\epsilon^{\mu\nu}\partial_\mu\Omega\,\partial_\nu P=0.
\label{GraFloEqu_1}
\end{alignat}
In the orthonormal complex coordinates introduced in
Sec.~\ref{s:bogomolny}, this equation can be written as
\begin{alignat}1
-\alpha[P,\partial\bar\partial P]+\beta\left(\partial\Omega\,\bar\partial P-\bar\partial\Omega\,\partial P\right)
=0 .
\label{GraFloEqu_2}
\end{alignat}

The first term is the contribution from the metric action \(S_g\) and drives 
the projector toward Bogomolny saturation. The second term is the contribution
from \(S_\Omega\) and drives the Berry curvature toward a uniform 
distribution. The no-go theorem above shows that, in a finite-dimensional
rank-one Chern band, these two requirements cannot in general be satisfied
simultaneously. Therefore, the fixed point of the combined flow is expected to
be a nontrivial balance between the two terms in Eqs.~\eqref{GraFloEqu_1} or
\eqref{GraFloEqu_2}, rather than an exactly ideal configuration.


\subsubsection{Lowest Landau level and ideal quantum geometry}\label{s:LLL}

The no-go theorem discussed above applies to finite-dimensional Bloch wave
functions. It does not exclude ideal quantum geometry in an infinite-dimensional
Hilbert space. The lowest Landau level provides a canonical example of such an
ideal geometry: it satisfies the Bogomolny condition and has uniform Berry
curvature
\cite{Ozawa:2021vs,PhysRevB.104.045104,1zg9-qbd6}.

To discuss it more precisely,  we consider a model 
defined in the continuum space described by coordinates $\bm x$.
Consider a charged particle of charge $e$ on a two-dimensional plane in a uniform magnetic field $B$ 
perpendicular to the plane. We introduce orthonormal real-space coordinates $y_a$ 
by $\bm x = \bm e^ay_a$ and denote real-space 
derivatives by
\begin{alignat}1
\bm\nabla^a=\frac{\partial}{\partial y_a}. 
\end{alignat}
This notation is used to distinguish them from momentum-
space derivatives. The covariant derivative is
\begin{alignat}1
D^a=\bm\nabla^a-ieA^a . 
\end{alignat}
Introducing complex coordinates
\begin{alignat}1
z=y_1+iy_2,
\quad
\bar z=y_1-iy_2,
\label{ComCoo}
\end{alignat}
we write the corresponding covariant derivatives as \(D\) and \(\bar D\). The
lowest-Landau-level condition is
\begin{alignat}1
D\Psi=0
\quad \text{or} \quad
\bar D\Psi=0,
\end{alignat}
depending on the sign of \(eB\).

We now impose magnetic translational symmetry towards the directions of $\bm b^\mu$, which are
dual vectors of the reciprocal vectors $\bm b_\mu$ with (\ref{DuaBas}).
To this end, let us introduce the coordinate $x_\mu$ to these directions and write
$\bm x=\bm b^\mu x_\mu$.
For a generic magnetic field, ordinary translational symmetry is replaced by
magnetic translation symmetry. A magnetic unit cell of area \({\cal A}\) is
compatible with the magnetic translations when the flux through the unit cell is
quantized as
\begin{alignat}1
e\Phi=-eB{\cal A}= 2\pi Q, \quad
(Q\in\mathbb Z).
\end{alignat}
With the present sign convention, the lowest Landau level considered below has
\(C=-1\) for one flux quantum, corresponding to \(Q=-1\).
In what follows, we restrict our discussions to this case.
The complex coordinate (\ref{ComCoo}) can be written as
\begin{alignat}1
z=e^\mu x_\mu,\quad
\bar z=\bar e^\mu x_\mu,
\end{alignat}
where $e^\mu$ is defined using Eq. (\ref{VieBei}) by
\begin{alignat}1
e^\mu=e_1{}^\mu+i e_2{}^\mu,\quad
\bar e^\mu=e_1{}^\mu-i e_2{}^\mu .
\end{alignat}
A translation along the \(\mu\) direction therefore acts as 
\begin{alignat}1
z\mapsto z+e^\mu,
\qquad
\bar z\mapsto \bar z+\bar e^\mu .
\end{alignat}

It is convenient to introduce
\begin{alignat}1
w=\frac{z}{e^1},\quad
\tau=\frac{e^2}{e^1},
\end{alignat}
so that \(\tau\) is the modular parameter of the torus. For one flux quantum,
an unnormalized lowest-Landau-level state can be written in terms of a theta function as
\cite{PhysRevResearch.6.033238}
\begin{alignat}1
\braket{\bm x}{W}=
e^{-\frac{\pi}{2\tau_2}
\left(|w|^2-w^2\right)
+\frac{\pi\kappa(w-\bar w)}{\tau_2}}
\vartheta(w+\kappa,\tau),
\end{alignat}
where $\tau_2=\Im \tau$ and
\begin{alignat}1
\kappa=\frac{\tau k^1-k^2}{2\pi}.
\end{alignat}
The important point is that this wave function depends on \(\kappa\), but not on \(\bar\kappa\).
Thus, as a Bloch frame over the momentum-space torus, it
is holomorphic in \(\kappa\). This is the Landau-level realization of the
Bogomolny condition.

The infinite-dimensional nature of this construction is manifest if we expand
\(\ket{W}\) in a complete real-space basis. Let
\(\braket{\bm x}{\phi_n}=\phi_n(w,\bar w)\), \(n=0,1,\ldots\), be a complete
orthonormal set. Then
\begin{alignat}1
\braket{\bm x}{W}=\sum_{n=0}^{\infty}\phi_n(w,\bar w) W_n(\kappa),
\end{alignat}
where $W_n(\kappa)=\braket{\phi_n}{W}$.
Thus the Bloch frame is an infinite-dimensional vector,
\begin{alignat}1
W(\kappa)=\left( W_0(\kappa), W_1(\kappa), \ldots \right)^T .
\end{alignat}
This is precisely the possibility excluded in the finite-dimensional no-go theorem.

For the above state, the Hermitian norm is
\begin{alignat}1
H&=\braket{W}{W}=\int_0^1 dx_1\int_0^1 dx_2\left|\braket{\bm x}{W}\right|^2
\nonumber\\
&=\frac{1}{\sqrt{2\tau_2}}e^{-\frac{\pi(\kappa-\bar\kappa)^2}{2\tau_2}}.
\end{alignat}
Therefore, the Berry curvature on the \(\kappa\) torus is 
\begin{alignat}1
\Omega_\kappa=-2\partial_{\bar\kappa}\partial_\kappa \ln H=-\frac{2\pi}{\tau_2}. 
\end{alignat}
It is manifestly uniform. The volume element on the \(\kappa\) torus is 
\begin{alignat}1
d\mu_\kappa=\frac{i}{2}d\kappa\wedge d\bar\kappa=\frac{\tau_2}{(2\pi)^2}dk^1\wedge dk^2 .
\end{alignat}
Hence
\begin{alignat}1
\Omega_\kappa d\mu_\kappa=-\frac{1}{2\pi}dk^1\wedge dk^2 .
\end{alignat}
Integrating over \(0\le k^1,k^2<2\pi\), we obtain
$C=-1 $.
Thus the lowest Landau level satisfies both the Bogomolny condition and the
uniform-curvature condition. It realizes the formal lower bound not by a
finite-dimensional Bloch frame, but by an infinite-dimensional
Hilbert-space-valued holomorphic frame.

\section{Numerical results for lattice models}\label{s:numerics}

In this section, we study the gradient flow for lattice Chern bands.
As discussed above, Bogomolny saturation and uniform Berry curvature
cannot, in general, be realized simultaneously for finite-dimensional
projectors.  We therefore study how the metric term and the
Berry-curvature term improve these two properties in concrete lattice
models.  We first consider the Wilson-Dirac model and then the 
Hofstadter model.

\subsection{Wilson-Dirac model} 

As an example, we consider the Wilson-Dirac model, which was also studied 
for the metric gradient flow in Ref.~\cite{5vhk-7x54}. The Hamiltonian is defined by
\begin{alignat}1
H&=\frac{it}{2}\sum_j\Big(\bm c_{j+\hat x}^\dagger\sigma_1\bm c_j+\bm c_{j+\hat y}^\dagger \sigma_2\bm c_j\Big)
+\mbox{H.c.}+m\sum_j\bm c_j^\dagger \sigma_3\bm c_j
\nonumber\\
&+\frac{b}{2}\sum_j\sum_{\mu=x,y}
\Big(\bm c_{j+\hat \mu}^\dagger \sigma_3\bm c_j+\bm c_{j}^\dagger \sigma_3\bm c_{j+\hat \mu}
-2\bm c_j^\dagger \sigma_3\bm c_j\Big),
\end{alignat}
where $\bm c_j^\dagger=(c_{j1}^\dagger,c_{j2}^\dagger)$ are two-component creation operators. 
The corresponding Bloch Hamiltonian $H=\sum_k\bm c_k^\dagger {\cal H}(k)\bm c_k$ is given by
\begin{alignat}1
{\cal H}(k)=t\sum_{\mu=1,2}\sigma_\mu\sin k^\mu
+\sigma_3\left[m+b\sum_{\mu=1,2}(\cos k^\mu-1)\right],
\end{alignat}
where \(\sigma_\mu\) are the Pauli matrices. For \(t=b=1\), the occupied band
has Chern number \(C=-1\) for \(2<m<4\), \(C=+1\) for \(0<m<2\), and \(C=0\)
otherwise. In the following, we take \(m=2.2\), so that the initial occupied
band has \(C=-1\). 
We construct the initial projector \(P(k)\) from the occupied
eigenstate of \(H(k)\) and evolve it according to the gradient-flow equations 
\eqref{GraFloEqu} and \eqref{J}.

\begin{figure*}[thb]
\centering
\includegraphics[width=0.99\linewidth]{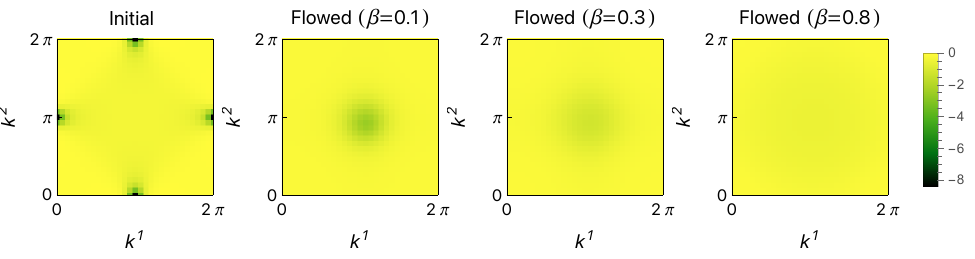}
\caption{
Berry curvature \(\Omega(k)\) for the Wilson-Dirac model with mass parameter 
\(m=2.2\) and \(t=b=1\), computed on a \(31^2\) mesh in the Brillouin zone.
The leftmost panel labeled ``Initial'' is computed from the initial projector
\(P(k)\), while the other panels labeled ``Flowed'' are computed from the final
projectors obtained by the gradient flow with
\((\alpha,\beta)=(0.9,0.1)\), \((0.7,0.3)\), and \((0.2,0.8)\), respectively.
}
\label{f:omega}
\end{figure*}

\subsubsection{Fixed points of the gradient flow}
Figure~\ref{f:omega} shows the Berry-curvature distribution, computed using the 
plaquette formulation of \cite{FHS05}, before and after
the gradient flow for several choices of the weights \((\alpha,\beta)\). 
Here, the Berry curvature $\Omega(k)$ is plotted as the plaquette Berry flux 
$\Phi(k)=-\Im\ln\left[U_1(k)U_2(k+\hat1)U_1^{-1}(k+\hat2)U_2^{-1}(k)\right]$ 
divided by the plaquette area $d\mu_h$, \(\Omega(k)=\Phi(k)/d\mu_h\), 
where $U_\mu(k)=\Det_{i,j}\braket{\Psi_i(k)}{\Psi_j(k+\hat\mu)}$ and $\hat \mu$ denotes the 
unit vector in the $\mu$ direction on the discretized Brillouin zone.
The initial Wilson-Dirac state has a strongly nonuniform Berry curvature, with a 
pronounced negative peak in the Brillouin zone around $(k_1,k_2)=(\pi,0)$ and $(0,\pi)$. 
As demonstrated in Ref. \cite{5vhk-7x54}, the quantum-metric gradient flow ($\beta=0$) 
makes the peak of the Berry curvature move around $(\pi,\pi)$.
When the metric term is dominant,
as in \((\alpha,\beta)=(0.9,0.1)\), the final state still exhibits a localized
structure in \(\Omega(k)\). As the weight of the curvature term is increased,
the Berry curvature becomes progressively smoother. In particular, for
\((\alpha,\beta)=(0.1,0.9)\), the final configuration has an almost uniform
Berry curvature over the Brillouin zone.

To quantify the curvature fluctuation, we define
\begin{alignat}1
\delta_\Omega=\int_{T^2}\left(\Omega-\bar\Omega\right)^2\,d\mu,
\end{alignat}
where $\bar\Omega$ is the mean value of the Berry curvature (\ref{MeaOme}). 
This quantity measures the deviation of the Berry curvature from the uniform
distribution. To compare the effect of the flow for different choices of
\((\alpha,\beta)\), we normalize it by its initial value and define
\begin{alignat}1
R_\Omega=\frac{\delta_\Omega^{\rm final}}{\delta_\Omega^{\rm initial}} .
\label{RO}
\end{alignat}
Thus \(R_\Omega<1\) indicates that the Berry-curvature fluctuation is reduced 
by the gradient flow.

We next examine the deviation from the Bogomolny condition. Since the Chern
number is negative in the present case, the Bogomolny condition is
$\bar\partial P P=0 $.
We define the Bogomolny defect as
\begin{alignat}1
\delta_{\rm B}=
\int_{T^2}\|\bar\partial P P\|^2\,d\mu ,
\end{alignat}
and further, the normalized Bogomolny defect by
\begin{alignat}1
R_{\rm B}=\frac{\delta_{\rm B}^{\rm final}}{\delta_{\rm B}^{\rm initial}} .
\label{RB}
\end{alignat}
Again, \(R_{\rm B}<1\) indicates that the Bogomolny defect is reduced by the
gradient flow.

\begin{figure}[thb]
\centering
\includegraphics[width=0.99\linewidth]{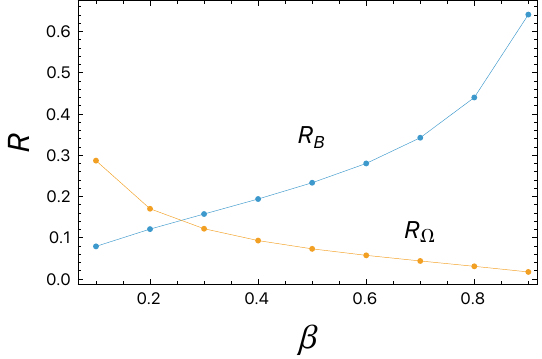}
\caption{
Dependence of the ratios \(R_\Omega\) and \(R_{\rm B}\), defined in
Eqs.~(\ref{RO}) and \eqref{RB}, on the relative weight \(\beta\), with
\(\alpha+\beta=1\).  The two ratios exhibit opposite trends as \(\beta\)
is increased.
}
\label{f:R}
\end{figure}

The numerical values of these normalized defects are summarized in
Fig. \ref{f:R}. We find that the curvature fluctuation is most strongly
suppressed when \(\beta\) is large. In contrast, the Bogomolny defect is most
strongly reduced when \(\alpha\) is large. This is consistent with the fact that
the \(S_\Omega\) term drives the system toward uniform Berry curvature, whereas
the metric term \(S_g\) drives the projector toward the Bogomolny condition.

These results show a clear trade-off between the two geometric requirements.
The metric term favors Bogomolny saturation, whereas the curvature term favors
uniform Berry curvature. In the finite-dimensional lattice model, the gradient
flow therefore is expected to approach  a nontrivial fixed point that balances these two
effects, rather than to a configuration satisfying both conditions exactly.

\begin{figure*}[thb]
\centering
\begin{tabular}{cc}
\includegraphics[width=0.72\linewidth]{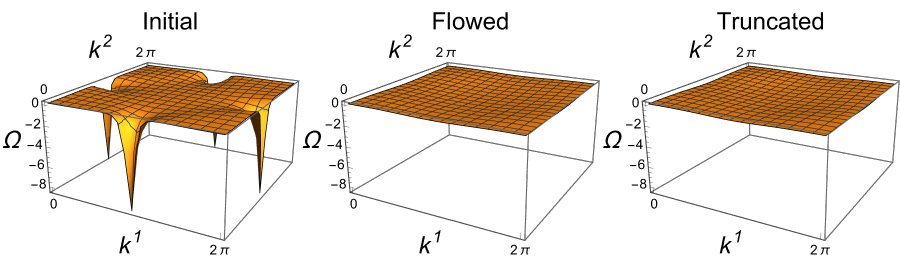}&
\includegraphics[width=0.28\linewidth]{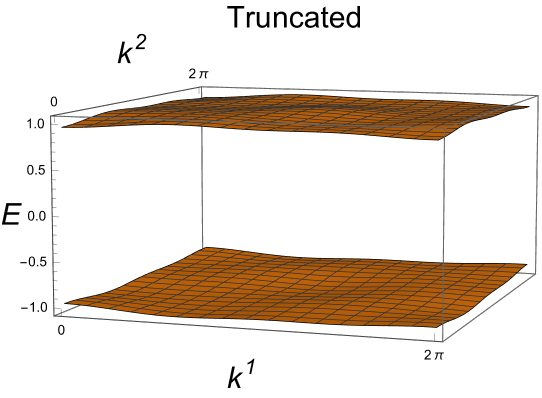}
\end{tabular}
\caption{
Three left panels show the Berry curvature \(\Omega(k)\) for the initial, 
flowed, and truncated models, as three-dimensional surface plots, complementary
to the density plots in Fig.~\ref{f:omega}.
The parameters are the same as those in
Fig.~\ref{f:omega}, except that \((\alpha,\beta)=(0.3,0.7)\). For the truncated model,
the range of $\Omega$ is $\Delta\Omega=0.50$.
The rightmost
panel shows the energy spectrum of the truncated Hamiltonian in
Eq.~\eqref{TruHam} with the hopping matrices given in Eqs.~\eqref{HopParOn}--\eqref{HopParNex}.
The bandwidth of the truncated model is $\Delta E=0.08$.
}
\label{f:omega_0p3}
\end{figure*}

\subsubsection{Short-range truncated flattened Hamiltonian}
We next construct a short-range lattice Hamiltonian from the projector obtained
after the gradient flow. From a spectral projector \(P(k)\), we define the
flattened Hamiltonian 
\begin{alignat}1
Q(k)=1-2P(k).
\label{FlatHam}
\end{alignat}
This Hamiltonian has two perfectly flat bands at energies \(-1\) and \(+1\).
However, its Fourier transform generally contains long-range hopping terms \cite{Sun:2011aa,Neupert:2011aa,Chen:2014aa,PhysRevB.104.115160}. 
We therefore truncate \cite{PhysRevB.104.115160} the real-space hopping amplitudes of \(Q(k)\), retaining
only the onsite term, the nearest-neighbor hoppings with displacements
\(\hat d=\pm\hat x\) and \(\pm\hat y\), the diagonal next-nearest-neighbor hoppings with
displacements \(\hat d=\pm\hat x\pm\hat y\), 
where $\hat x=(1,0)$ and $\hat y=(0,1)$.
We refer to the resulting model as the short-range truncated flattened Hamiltonian.
Explicitly, the truncated Hamiltonian in the second-quantized form is written as
\begin{alignat}1
H_{\rm tr}=&
\sum_j\sum_{\hat d} \bm c_{j+\hat d}^\dagger t_{\hat d}\bm c_j.
\label{TruHam}
\end{alignat}

We now construct a flattened Hamiltonian from the flowed projector.  In
the following, we focus on the flowed state obtained for
\((\alpha,\beta)=(0.3,0.7)\).  This choice corresponds to a representative
near-stationary state of the gradient flow for which the Berry curvature
is substantially flattened, while the projector remains sufficiently
short-ranged.  As a result, after truncation, both the band flatness and
the nearly uniform Berry curvature are well preserved.  For this flowed
projector, the dominant hopping matrices of the short-range truncated
flattened Hamiltonian are obtained as follows:
the onsite term is 
\begin{alignat}1
t_{\hat 0}
&=\begin{pmatrix}
0.48 & 0.03-0.03 i \\
0.03+0.03 i & -0.48
\end{pmatrix}.
\label{HopParOn}
\end{alignat}
The nearest-neighbor hopping matrices are
\begin{alignat}1
t_{\hat x}
&=\begin{pmatrix}
0.25-0.01 i & -0.02-0.32 i \\
-0.01-0.3 i & -0.25+0.01 i
\end{pmatrix},
\nonumber\\
t_{\hat y}
&=\begin{pmatrix}
0.25-0.01 i & -0.3+0.01 i \\
0.32-0.02 i & -0.25+0.01 i
\end{pmatrix}.
\label{HopParNea}
\end{alignat}
The diagonal next-nearest-neighbor hopping matrices are  
\begin{alignat}1
t_{\hat x+\hat y}
&=\begin{pmatrix}
-0.13+0.01 i & 0.1+0.09 i \\
-0.09+0.1 i & 0.13-0.01 i
\end{pmatrix},
\nonumber\\
t_{\hat x-\hat y}
&=\begin{pmatrix}
-0.13 & -0.1+0.1 i \\
0.09+0.09 i & 0.13
\end{pmatrix}.
\label{HopParNex}
\end{alignat}
Of course, $t_{-\hat d}=t_{\hat d}^\dagger$ in Eq. (\ref{TruHam}).
We note that this truncation is not an arbitrary choice of hopping range.  If the
hopping matrices in the Fourier expansion of the flattened Hamiltonian are
ordered by their norms,
\(\norm{t_{\hat d}}^2=\Tr t_{\hat d}^\dagger t_{\hat d}\), 
the dominant terms are precisely those retained above.  
\cin{The remaining hopping
matrices have smaller norms and give longer-range corrections to the exactly
flattened Hamiltonian.  Thus, for the present flowed projector, the Fourier
components are already dominated by short-range hoppings.  The gradient flow
also changes the complex phases of the hopping matrices, as seen in
Eqs.~\eqref{HopParOn}--\eqref{HopParNex}.  These phases are not put in by 
hand, but are generated by the flowed projector and reflect the redistribution
of the Berry curvature in momentum space.  The truncated model is therefore
obtained by keeping the largest hopping amplitudes of the flattened
Hamiltonian, including their complex phases.}

Figure~\ref{f:omega_0p3} compares the Berry curvature of the initial model, the
flowed model, and the truncated model using three-dimensional surface plots.
This representation is complementary to the density plots in
Fig.~\ref{f:omega} and makes the smoothing of the Berry-curvature profile more
visually transparent.

The initial Wilson-Dirac model has a 
strongly nonuniform Berry curvature. The gradient flow smooths the Berry
curvature profile, producing a nearly uniform distribution. Although the
corresponding flattened Hamiltonian contains long-range hopping terms, the
short-range truncation preserves the almost uniform Berry curvature remarkably
well. Moreover, the energy spectrum of the truncated Hamiltonian remains nearly
flat, as shown in the rightmost panel of Fig.~\ref{f:omega_0p3}. Thus the 
gradient-flow construction provides a practical route to a short-range lattice
Hamiltonian whose band is nearly flat and whose Berry curvature is nearly
uniform.

\subsection{Hofstadter model}
\label{subsec:hofstadter}

We finally apply the gradient flow to the Hofstadter model on the square
lattice.  We consider a magnetic flux
$  \phi=p/q$ per plaquette.  In the Landau gauge, the magnetic unit cell is enlarged
in the \(x\) direction and contains \(q\) sites, i.e., $\bm b^1=q\tilde{\bm b}^1$, 
where $\tilde{\bm b}^1$ stands for the original unit vector for the 1-direction of the square lattice.   
In accordance with the
notation used above, we denote by \(\bm b_\mu\) the reciprocal basis
associated with this magnetic unit cell, whereas we denote by
\(\tilde{\bm b}_\mu\) the reciprocal basis of the original microscopic
square lattice.  Since the magnetic unit cell is \(q\) times larger in
the \(x\) direction, as mentioned above, the two reciprocal bases are related by 
\begin{alignat}{1}
  \bm b_1=\frac{1}{q}\tilde{\bm b}_1,
  \qquad
  \bm b_2=\tilde{\bm b}_2 .
\end{alignat}
Thus the Brillouin zone associated with the magnetic unit cell is
reduced by a factor of \(q\) in the \(x\) direction. 

Let \(k^\mu\) be the momentum coordinates with respect to the magnetic
reciprocal basis \(\bm b_\mu\), while \(\tilde{k}^\mu\) denotes the
coordinates with respect to the microscopic reciprocal basis
\(\tilde{\bm b}_\mu\).  The same momentum vector can be written as
\begin{alignat}{1}
  \bm k=k^\mu \bm b_\mu=\tilde{k}^\mu \tilde{\bm b}_\mu .
\end{alignat}
Using \(\bm b_1=\tilde{\bm b}_1/q\) and \(\bm b_2=\tilde{\bm b}_2\), we obtain
\begin{alignat}{1}
  k^1=q\tilde{k}^1,
  \qquad
  k^2=\tilde{k}^2 .
\end{alignat}
By definition,  \(k^1\) has period \(2\pi\),  $0\leq k^1<2\pi$, while the microscopic momentum
\(\tilde{k}^1\) ranges over the reduced interval,
\begin{alignat}{1}
  0\leq \tilde{k}^1<\frac{2\pi}{q}.
\end{alignat}
In the numerical calculation below, we use the microscopic coordinate
\(\tilde{k}^1\).  
The Hamiltonian, however, depends on the magnetic-cell momentum
\(k^1=q\tilde{k}^1\).

Let us now discuss the role of the metric \(h_{\mu\nu}\).  
For the square lattice, we normalize the microscopic reciprocal basis so that
$\tilde{\bm b}_\mu\cdot\tilde{\bm b}_\nu= \delta_{\mu\nu}$.
Then, 
the metric in the magnetic-cell coordinates
is $h_{\mu\nu}=\mbox{diag}(1/q^2,1)$ ($\sqrt{h}=1/q$).
This shows why the metric formulation is useful.  The functional is
written in a coordinate-independent form, so that the same physical
quantity can be evaluated either in the magnetic-cell coordinates
\(k^\mu\), where the metric is nontrivial, or in the microscopic
coordinates \(\tilde{k}^\mu\), where the metric is flat.

\begin{figure}[htb]
  \centering
  \includegraphics[width=\linewidth]{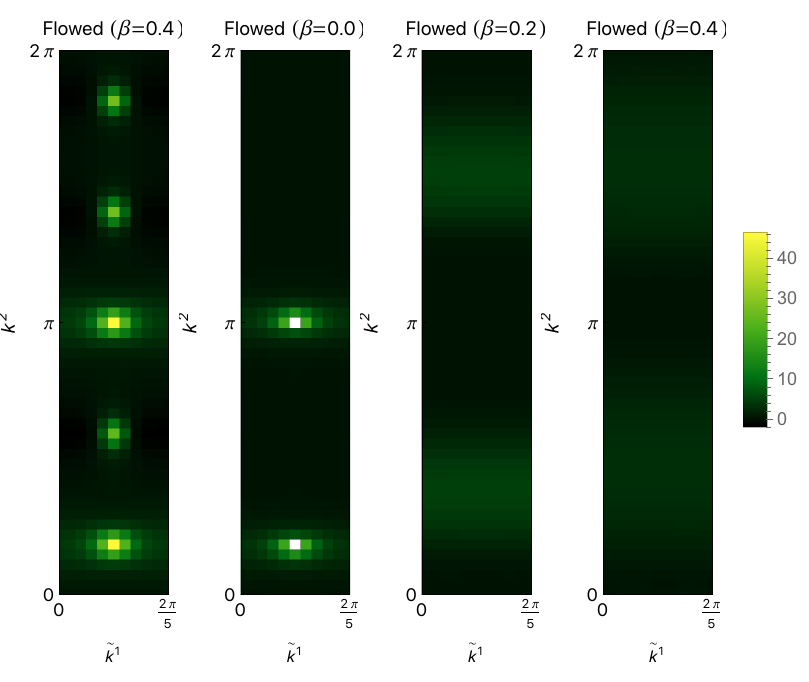}
  \caption{
  Berry curvature distribution for the lowest Hofstadter band at
  \(\phi=2/5\), with Chern number \(C=2\).  The horizontal axis is the
  microscopic momentum \(\tilde{k}^1\).  The panels show the initial
  projector and flowed projectors for
  \((\alpha,\beta)=(1.0,0.0)\), \((0.8,0.2)\), and \((0.6,0.4)\), with
  \(\alpha+\beta=1\).  Increasing \(\beta\) suppresses the curvature
  peaks and makes the Berry curvature more uniform.  The color scale is
  chosen to resolve the initial distribution; in the
  \((\alpha,\beta)=(1.0,0.0)\) panel, the Berry curvature has two
  sharply localized peaks whose maximum value is approximately
  \(\Omega\simeq 240\), outside the displayed color range.
  }
  \label{fig:hofstadter_omega}
\end{figure}

Below we consider the case
\(\phi=2/5\).  For this flux, the lowest band has Chern number \(C=2\).
Figure~\ref{fig:hofstadter_omega} shows the Berry curvature distribution
of the lowest band before and after the flow for several choices of
\((\alpha,\beta)\), with \(\alpha+\beta=1\).  For
\((\alpha,\beta)=(1,0)\), the Berry curvature is concentrated into two
sharply localized peaks.  This is consistent with the general behavior
of the metric-dominant flow, which tends to produce \(C\) localized
curvature peaks for a band with Chern number \(C\); in the Wilson-Dirac 
model with \(C=-1\), this appears as a single negative peak.  Introducing a finite
\(\beta\) suppresses these peaks and drives the Berry curvature toward a
more uniform distribution over the Brillouin zone.

\section{Conclusion}\label{s:conclusion}

In this paper, we proposed a gradient-flow approach to quantum states with
ideal quantum geometry. The action consists of the quantum-metric term and the
square of the Berry curvature. The resulting gradient-flow equation contains
two competing effects: the metric term drives the projector toward the
Bogomolny bound, whereas the Berry-curvature term suppresses the spatial
fluctuation of the Berry curvature and drives the system toward uniform
curvature.

The fixed point of the gradient flow is nontrivial. For finite-dimensional
Bloch wave functions, such as those in lattice models, the Bogomolny bound and
the exactly uniform Berry-curvature condition cannot in general be satisfied
simultaneously. In contrast, the lowest Landau level in the continuum realizes
both conditions, because it is described by an infinite-dimensional
Hilbert-space-valued Bloch frame. Thus the present analysis clarifies the 
difference between ideal quantum geometry in the continuum and its realization
in finite-dimensional lattice models.

We also examined the proposed gradient flow numerically in the
Wilson-Dirac model and the Hofstadter model.  Starting from the initial 
projector, we obtained flowed projectors in which the quantum geometry is
substantially improved.  In the Wilson-Dirac model, the Berry curvature 
becomes smoother as the weight of the Berry-curvature term is increased,
while the Bogomolny defect is more strongly suppressed when the metric
term is dominant.  This demonstrates the trade-off between Bogomolny
saturation and uniform Berry curvature in finite-dimensional lattice
models.  We further confirmed the same tendency in the Hofstadter model:
for a Chern band with \(C=2\), the metric-dominant flow produces two
localized extrema of the Berry curvature, while the Berry-curvature term
suppresses these extrema and drives the curvature toward a more uniform
distribution.

Finally, we constructed a short-range truncated flattened Hamiltonian
from the flowed projector.  Although the exactly flattened Hamiltonian
generally contains long-range hopping terms, we found that truncating the
hopping amplitudes up to next-nearest-neighbor terms can still produce a
lattice model with nearly flat bands and nearly uniform Berry curvature.
This suggests that the present gradient-flow approach provides a
practical route to constructing short-range lattice models with almost
ideal quantum geometry.

Although the formulation developed in this paper applies to projectors of
arbitrary rank, our numerical examples focused mainly on rank-one Chern bands.
A natural future direction is to make the present gradient-flow approach a more
systematic method for designing lattice models with nearly ideal quantum
geometry.  In this work, the flattened Hamiltonian \(Q=1-2P\) collapses the
unoccupied subspace into a degenerate sector.  It would be interesting to go
beyond this simple form and construct more general and natural lattice models with several
nondegenerate bands, keeping nearly ideal geometry for selected bands.

\cin{We have considered the flow for projectors in lattice models.  In this
sense, the present method can be applied to any isolated band in a lattice
Hamiltonian.  On the other hand, higher Landau levels in the continuum have a
different geometric relation, such as
\((2n+1)|C|\) in the integrated trace condition.  Such a relation reflects
additional Landau-level structure and is not expected to be preserved by the
unconstrained projector flow used here.  It would be interesting to construct
a constrained flow, or a different geometric functional, which is adapted to
higher-Landau-level-type quantum geometry.}

\acknowledgements
This work was supported in part by a Grant-in-Aid for Scientific Research
(Grant No.~26K06957) from the Japan Society for the Promotion of Science.

\bibliography{uniform_F_gf}


\end{document}